\documentclass[amsfonts]{article}
\usepackage{graphicx}

\newcommand{\ket}[1]{{|#1 \rangle}}

\newcommand{\kpsi}{\ket{\psi}}
\newcommand{\kPsi}{\ket{\Psi}}

\newcommand{\expect}[1]{{\langle #1 \rangle}}
\newcommand{\adj}[1]{{#1}^\dagger}

\newcommand{\reg}[1]{\mathbf{#1}}

\newcommand{\mod}{\,{\rm mod}\,}
\newcommand{\hilb}{\mathcal{H}}
\newcommand{\qubit}{\mathcal{B}}
\newcommand{\compl}{\mathbf{C}}

\newcommand{\integ}{\mathbf{Z}}

\newcommand{\mwith}{\quad\mathrm{with}\quad}

\newcommand{\lab}[1]{\label{#1}}

\newcommand{\tab}[3]{
  \begin{center}\small
    \begin{tabular}{#1}
      \hline
      #2 \\\hline\hline 
      #3
    \end{tabular}
  \end{center}
}

\newcommand{\fref}[1]{figure~\ref{#1}}
\newcommand{\Fref}[1]{Figure~\ref{#1}}
\newcommand{\eqref}[1]{(\ref{#1})}

\newcommand{\matr}[2]{
  \left(\begin{array}{#1}
  #2
  \end{array}\right)
}
\newcommand{\alt}[1]{
  \left\{ \begin{array}{cl}#1\end{array}\right.
}

\newcommand{\graphic}[3]{
  \begin{figure}[h]
  \begin{center}
  \includegraphics[width=#2]{#1.eps}
  \caption{\it #3}\lab{#1}
  \end{center}
  \end{figure}
}

\newcommand{\bqcl}{
  \begin{center}
  \sloppy\footnotesize
  \begin{minipage}{0.85\textwidth}
}
\newcommand{\eqcl}{
  \end{minipage}
  \end{center}
}

\newtheorem{definition}{Definition}

\begin{document}

\title{Classical Concepts in Quantum Programming}
\author{Bernhard \"Omer}
\date{November 11, 2002}


\begin{center}
~ \\
\vspace{5mm}
{\LARGE Classical Concepts in Quantum Programming}\\
\vspace{6mm}
{\large Bernhard \"Omer}\\
\vspace{6mm}
{Institute for Theoretical Physics}\\
{Technical University Vienna, Austria}\\
{\verb=oemer@tph.tuwien.ac.at=}\\
\vspace{6mm}
November 11, 2002
\vspace{6mm}
\end{center}

\begin{abstract}
The rapid progress of computer technology has been accompanied by a corresponding evolution of software development, from hardwired components and binary machine code to high level programming
languages, which allowed to master the increasing hardware complexity and fully exploit its potential. 

This paper investigates, how classical concepts like hardware abstraction, hierarchical programs, data types, memory management, flow of control and structured programming can be used in quantum computing. The experimental language QCL will be introduced as an example, how elements like irreversible functions, local variables and conditional branching, which have no direct quantum counterparts, can be implemented, and how non-classical features like the reversibility of unitary transformation or the non-observability of quantum states can be accounted for within the framework of a procedural programming language. 
\end{abstract}

\section{Quantum Programming}

\subsection{Quantum Programming Languages}

From a software engineering point of view, we can regard the 
formalism of Hilbert-space algebra as a specification language, as the mathematical description of a quantum algorithm is inherently declarative and provides no means to derive a unique decomposition into elementary operations for a given quantum hardware.

Low level formalisms like quantum circuits \cite{deutsch2}, on the other hand, are usually restricted to specific tasks, such as the
description of unitary transformations, and thus lack the 
generality to express all aspects of non-classical algorithms.

The purpose of programming languages is therefore twofold, as
they allow to express the semantics of the computation in 
an abstract manner, as well as the automated 
generation of a sequence of elementary operations to control
the computing device. 

\subsection{Quantum Algorithms}\lab{pqp-alg}

In its simplest form, a quantum algorithm merely consists
of a unitary transformation and a subsequent measurement
of the resulting state. For more ``traditional'' computational tasks,
however, as e.g. searching or mathematical calculations, efficient quantum implementations often have the form of probabilistic algorithms.
\Fref{alg} shows the basic outline of a probabilistic
non-classical algorithm with a simple evaluation loop.

\graphic{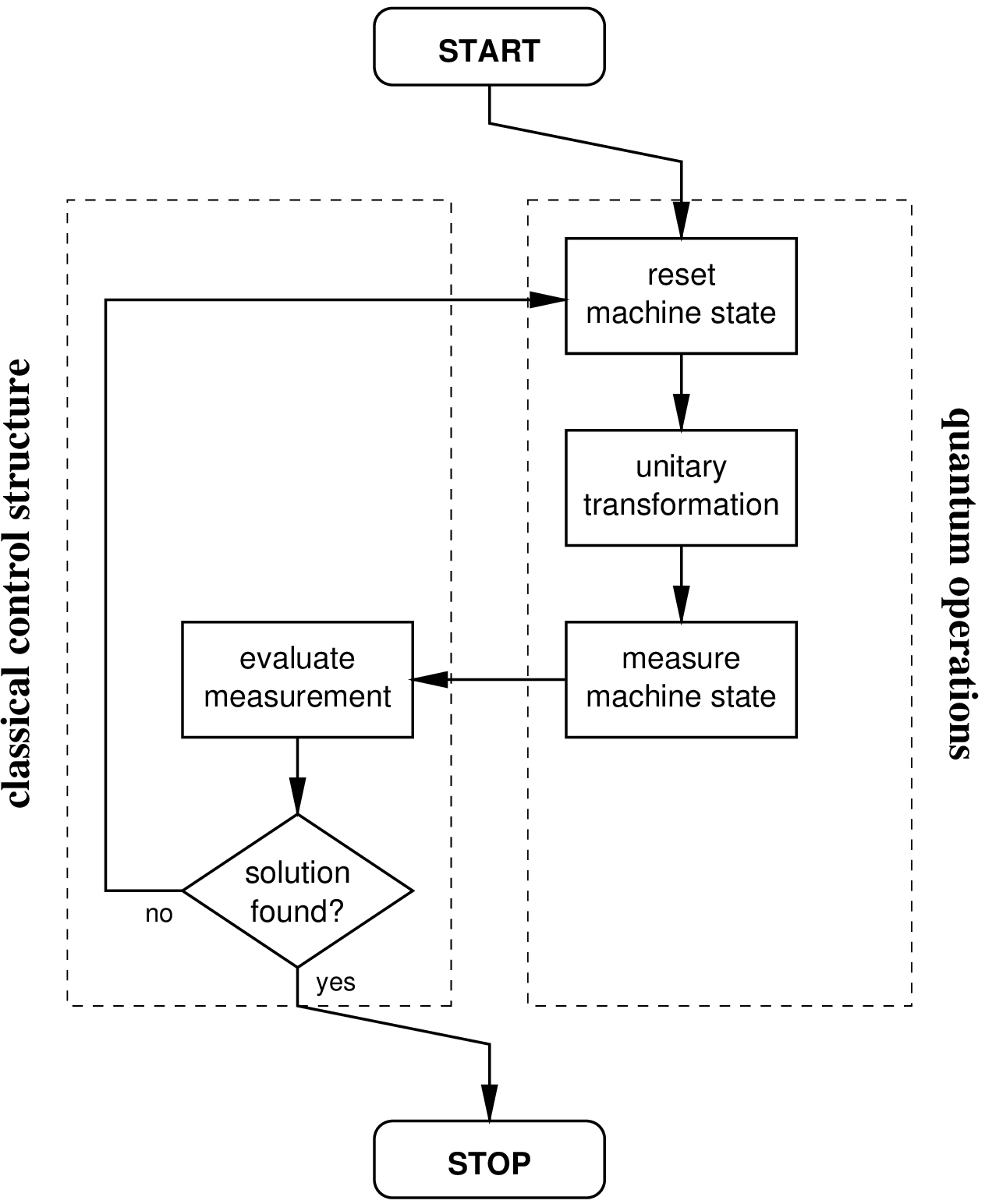}{60mm}{a simple non-classical algorithm}

More complex quantum algorithms, as e.g. Shor's algorithm for
quantum factoring \cite{shor,ekert}, can also include 
classical random numbers, partial measurements, nested
evaluation loops and multiple termination conditions; thus
the actual quantum operations are embedded into a classical 
flow-control framework.

In the discussion of non-classical algorithms, both aspects, the classical control structure and the actual quantum operations, are usually treated separately and often, only the latter is formally described \cite{wallace}.
In order to provide a consistent formalism, a quantum programming language will have to resolve this antagonism by generalizing existing classical programming concepts to the field of quantum computing.

\subsection{Structured Quantum Programming}

In traditional computing science, programming languages can be categorized as either logical (e.g. Prolog), functional (e.g. LISP) or procedural (e.g. Fortran, Pascal), the latter being the most widely 
used, for the description of algorithms, as well as the 
actual implementation of real world programs \cite{aaby}.

Procedural programming languages can be characterized by 

\begin{itemize}
\item {\bf explicit flow of control}
\item {\bf hierarchical program structure}
\item {\bf tight mapping between code and computation}
\end{itemize}

\noindent which seems to fit most people's way of reasoning
about computational tasks.
A procedural language is called {\it structured}, if flow control is restricted to selection- and loop-statements with well defined entry- and exit-points (e.g. Modula, Pascal without {\tt goto}-statement) \cite{dijkstra,dahl}.

Structured quantum programming is about extending these
concepts into the field of quantum computing while preserving their classical semantics.
The following table gives an overview of quantum language elements along with their classical counterparts.

\tab{|c|c|}{{\bf classical concept} & {\bf quantum analog}}{
classical machine model & hybrid quantum architecture \\\hline
variables & quantum registers \\\hline
subroutines & unitary operators \\\hline
argument and return types & quantum data types \\\hline
local variables & scratch registers \\\hline
dynamic memory & scratch space management \\\hline\hline
boolean expressions & quantum conditions \\\hline
conditional execution & conditional operators \\\hline
selection & quantum if-statement \\\hline
conditional loops & quantum forking \\\hline\hline
{\it none} & inverse execution of operators \\\hline
{\it none} & quantum measurement \\\hline
}

\subsection{Hybrid Architecture}\lab{pqp-arch}

Structured quantum programming uses a classical universal language to define the actual sequence of elementary instructions for a quantum computer, so a program is not intended to run on a quantum computer itself, but on a (probabilistic) classical computer, which in turn controls a quantum computer and processes the results of measurements.
In the terms of classical computer science, this architecture
can be described as a universal computer with a quantum oracle
(\fref{arch}).

\graphic{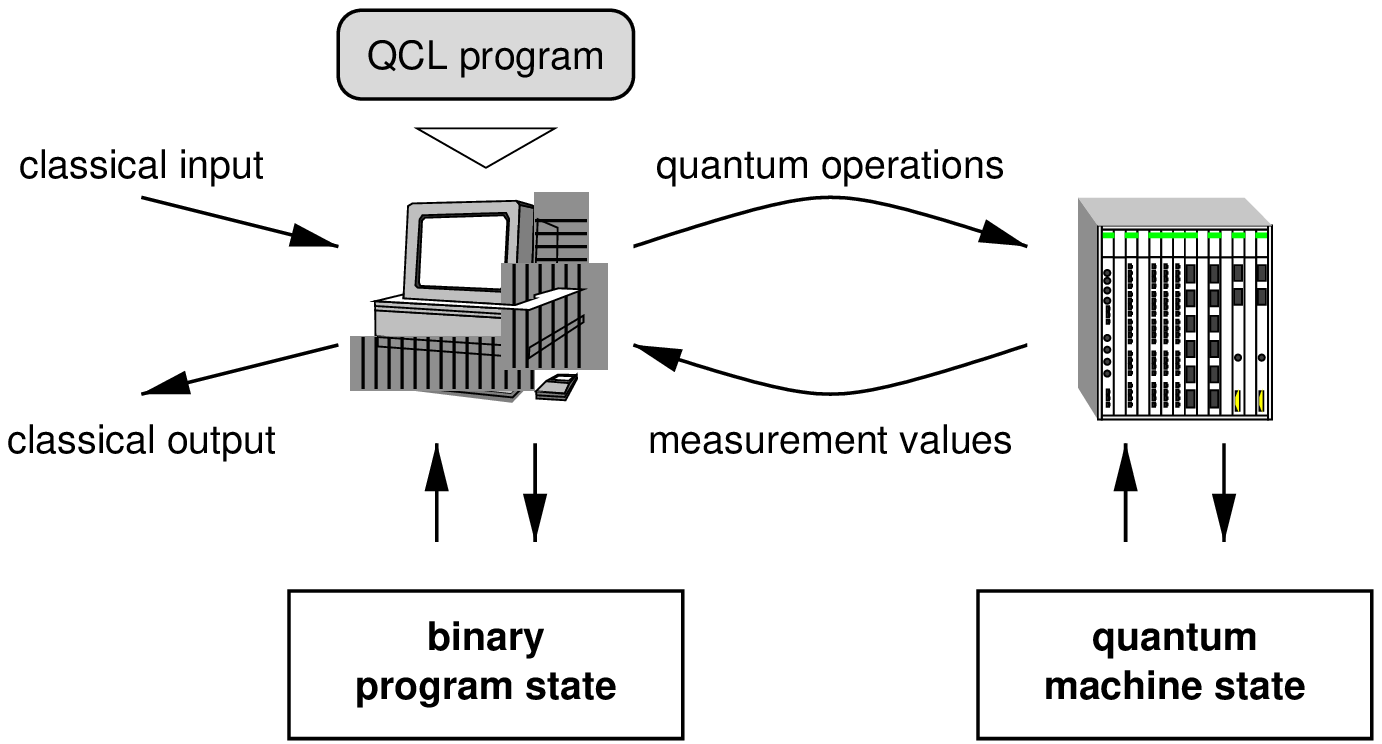}{80mm}{hybrid quantum architecture}

From the perspective of the user, a quantum program behaves exactly
like any other classical program, in the sense that it takes
classical input, such as startup parameters or interactive data,
and produces classical output.

The state of the controlling computer (i.e. 
program counter, variable values, but also the mapping of 
quantum registers) is referred to as {\it program state}.
The quantum computer itself does not require any control logic, its
computational state can therefore be fully described by the common 
quantum state $\kPsi$ of its qubits ({\it machine state}).

\section{The Programming Language QCL}\lab{qcl}

QCL (an acronym for ``quantum computation language'') is an 
experimental structured quantum programming language \cite{oemer1}.
A QCL interpreter, written in C++, including a numerical 
simulation library ({\tt libqc}) to emulate the quantum backend
is available from
\begin{quotation} 
{\tt http://tph.tuwien.ac.at/\verb=~=oemer/qcl.html}
\end{quotation}
\noindent as free software under the terms of the GPL.
 
\subsection{Quantum Storage and Registers}\lab{qcl-mem}

The smallest unit of quantum storage in QCL is the qubit

\begin{definition}[Qubit]
A qubit or quantum bit is a quantum system whose state $\kpsi\in\qubit$ can be fully described by a superposition of two orthonormal eigenstates labeled $\ket{0}$ and $\ket{1}$, i.e. $\qubit=\compl^2$.
\begin{equation} 
\ket{\psi}=\alpha \ket{0}+\beta \ket{1}
\mwith |\alpha|^2+|\beta|^2=1
\end{equation}
\end{definition}

QCL treats qubits as quantum registers of length 1.

\bqcl
\begin{verbatim}
qcl> qureg a[1]; qureg b[1]; // allocate 2 qubits
qcl> Rot(-pi/3,a);           // rotate 1st qubit
qcl> H(b);                   // Hadamard Transformation       
\end{verbatim}
\eqcl

\begin{definition}[Machine State]
The machine state $\kPsi\in\hilb$ of an $n$-qubit quantum computer is the state of a composite system of $n$ identical qubits, i.e. $\hilb=\qubit^{\otimes n}=\compl^{2^n}$.
\end{definition}

QCL --- if used together with a numerical simulator --- provides
debugging functions which allow the inspection of the otherwise
unobservable machine state.
In the example below  $\kPsi=\ket{00}\otimes(H\,\ket{0})\otimes R_x(\pi/3)\,\ket{0}$:

\bqcl
\begin{verbatim}
qcl> dump;                   // show product state as generated above
: STATE: 2 / 4 qubits allocated, 2 / 4 qubits free
0.612372 |0000> + 0.612372 |0010> + 0.353553 |0001> + 0.353553 |0011>
\end{verbatim}
\eqcl

\begin{definition}[Quantum Register]
An $m$ qubit quantum register $\reg{s}$ is a sequence of
mutually different qubit positions
$\expect{s_0, s_1 \ldots s_{m-1}}$ of some machine state 
$\ket{\Psi}\in\compl^{2^n}$ with $n\geq m$.

Using an arbitrary permutation $\pi$ over $n$ elements with
$\pi_i=s_i$ for $i<m$, a unitary reordering operator 
$\Pi_{\reg{s}}$ is defined as (see also \ref{qufunct})
\begin{equation}
  \Pi_{\reg{s}}\,\ket{d_0,d_1\ldots d_{n-1}}=
  \ket{d_{\pi_0},d_{\pi_1}\ldots d_{\pi_{n-1}}}
\end{equation}
\end{definition}

Quantum registers are the fundamental quantum data-type in QCL.
As they contain the mapping between the symbolic quantum
variables and the actual qubits in the quantum computer, they
are the primary interface between the classical frontend and
the quantum backend of the hardware architecture, as the machine 
state can only be accessed via registers.

Quantum registers are dynamically allocated and can be local.
Temporary registers can be created by using the qubit- 
({\tt q[n]}), subregister- ({\tt q[n:m]}) and 
concatenation-operators ({\tt q\&p}).

\subsection{Operators}\lab{qcl-ops}

\subsubsection{Register Operators}\lab{qcl-ops-reg}

\begin{definition}[Register Operator]
The register operator $U(\reg{s})$ for an $m$-qubit unitary 
operator $U:\compl^{2^m}\to\compl^{2^m}$ and an $m$-qubit
quantum register $\reg{s}$ on an $n$-qubit quantum computer
is the $n$-qubit operator
\begin{equation}
  U(\reg{s})=\adj{\Pi}_{\reg{s}}\,(U \otimes I(n-m))\,\Pi_s
\end{equation}
with an reordering operator $\Pi_{\reg{s}}$ and
the $k$-qubit identity operator $I(k)$.
\end{definition}

\graphic{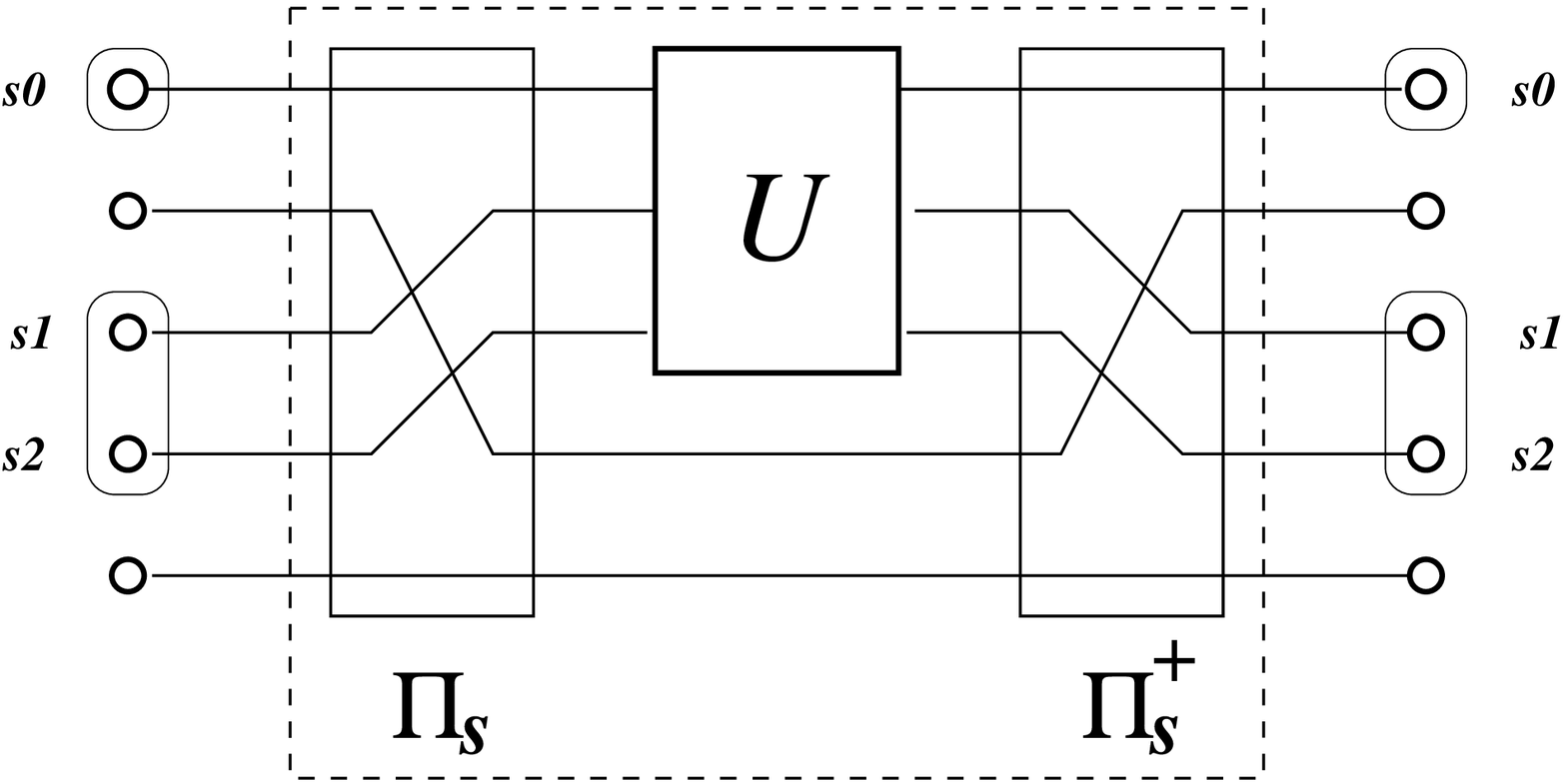}{60mm}{the register operator $U(\reg{s})$}

All operators in QCL are register operators and can also
have an arbitrary number of classical parameters.
The length of the operand-registers is passed as an additional 
implicit parameter.

\bqcl
\begin{verbatim}
operator dft(qureg q) { // Quantum Fourier Transform
  const n=#q;           // set n to length of input
  int i; int j;         // declare loop counters
  for i=1 to n {
    for j=1 to i-1 {    // apply conditional phase gates (see 2.3.2)
      if q[n-i] and q[n-j] { Phase(pi/2^(i-j)); }
    }
    H(q[n-i]);          // qubit rotation
  }    
  flip(q);              // swap bit order of the output
}
\end{verbatim}
\eqcl

QCL operators are unitary and have mathematical semantics i.e. their effect must be reproduceable and may only depend on the specified parameters. 
This esp. excludes

\begin{itemize}
\item dependencies on the program state (e.g. global variables)
\item side effects on the program state
\item user input and calls to {\tt random()}
\item non-unitary quantum operations (i.e. calls to {\tt measure}
  and {\tt reset})
\end{itemize}

For any QCL operator, the adjoint operator is determined on the
fly if the call is prefixed with the inversion-flag 
({\tt !}).\footnote{
Internally, this is achieved by recursively caching all 
suboperator calls and executing them in reverse order with
swapped inversion-flags.}

\bqcl
\begin{verbatim}
qcl> qureg q[2];             // allocate a 2-qubit register  
qcl> dft(q);                 // discrete Fourier transform
[2/32] 0.5 |00> + 0.5 |01> + 0.5 |10> + 0.5 |11>
qcl> !dft(q);                // inverse transform
[2/32] 1 |00>
\end{verbatim}
\eqcl

\subsubsection{Quantum Data Types}

Classical programming languages often allow to impose
access restrictions on variables and subroutine parameters.
This is equally done to prevent subsequent programming errors,
as to provide information to the compiler to allow for
more efficient optimizations.

QCL extends this concept to quantum registers by introducing
quantum data types to limit the ways how operators may effect
the machine state.

\tab{|l|l|}{{\bf type}&{\bf restriction}}{
{\tt qureg} & none \\
{\tt quconst} & invariant to all suboperators \\
{\tt quvoid} & has to be empty when the uninverted
  operator is called \\
{\tt quscratch} & has to be empty before and after the call \\\hline
}

\begin{definition}[Invariance of Registers]
A quantum register $\reg{c}$ is invariant
to a register operator $U(\reg{s},\reg{c})$ iff
$  U\,\ket{i}_\reg{s}\ket{j}_\reg{c}=
  \left(U_j\,\ket{i}_\reg{s}\right)\,\ket{j}_\reg{c}$ with unitary $U_j$.
\end{definition}
\begin{definition}[Empty Registers]
A register $\reg{e}$ is empty iff 
$\kPsi=\ket{0}_\reg{e}\ket{\psi'}$
\end{definition}

\subsubsection{Quantum Functions}\lab{qufunct}

One important aspect of quantum computing is, that --- due to the linearity of unitary
transformations --- an operator applied to a superposition state
$\kPsi$ is simultaneously applied to all basis vectors that
constitute $\kPsi$ ({\it quantum parallelism}) since
\begin{equation}
  U\,\sum_i c_i \ket{i}=\sum_i c_i (U\,\ket{i})
\end{equation}

In many cases $U$ implements a reversible boolean, or, 
equivalently, a bijective integer function, by treating the
basis vectors merely as bitstrings or binary numbers.

\begin{definition}
A $n$-qubit quantum function is a unitary operator 
of the form  $U:\ket{i}\to\ket{\pi_i}$ with some permutation 
$\pi$ over $\integ_{2^n}$.
\end{definition}

Quantum functions are implemented by the QCL subroutine-type {\tt qufunct}.

\bqcl\lab{inc}
\begin{verbatim}
qufunct inc(qureg x) {      // increment register
  int i;
  for i = #x-1 to 1 step -1 {
    CNot(x[i],x[0:i-1]);    // apply controlled-not from
  }                         // MSB to LSB
  Not(x[0]);
}
\end{verbatim}
\eqcl

To enforce the above restrictions, the 4 QCL subroutines types form a calling hierarchy, i.e. routines may only invoke subroutines of the same or a lower level.

\tab{|l|c|c|c|l|}{
{\bf subroutine} & $S$ & $\Psi$ & {\bf invertible} & {\bf description}
}{
{\tt procedure} & all & all & no & classical control structure\\
{\tt operator} & none & unitary & yes & general unitary operator\\
{\tt qufunct} & none & permutation & yes & quantum function \\
functions & none & none & no & mathematical functions
\\\hline
}

The columns $S$ and $\psi$ denote the allowed side-effects on the classical program state and the quantum machine state.

\subsubsection{Irreversible Functions}\lab{qcl-ops-qufunct}

One obvious problem in QC is its restriction to 
reversible computations.
Consider a simple operation like computing the parity of a bitstring
\begin{equation}
  \mathtt{parity}':\,\ket{i}\,\to\,\ket{b(i) \mod 2}
  \mwith b(n)=n\mod 2+b(\lfloor n/2 \rfloor),\, b(0)=0
\end{equation}
Clearly, this operation is non-reversible since 
$\mathtt{parity}'\,\ket{1}=\mathtt{parity}'\,\ket{2}$, so 
$\mathtt{parity}'$ is not an unitary operator.
However, if we use an additional target register, 
then we can define an operator $\mathtt{parity}$ which matches 
the condition $\mathtt{parity}\,\ket{i,0}=\ket{i,b(i)\mod 2}$.

\bqcl
\begin{verbatim}
qufunct parity(quconst x,quvoid y) {
  int i;
  for i = 0 to #x-1 {
    CNot(y,x[i]);           // flip parity for each set bit
  }
}
\end{verbatim}
\eqcl

In QCL, an operator $F: \ket{x}_\reg{x}\ket{0}_\reg{y} \to \ket{x}_\reg{x}\ket{f(x)}_\reg{y}$ is declared as {\tt qufunct} with at least one invariant ({\tt quconst}) argument 
register $\reg{x}$ and one empty ({\tt quvoid}) target register $\reg{y}$ as parameter.
The result for $F \ket{x}_\reg{x}\ket{y\ne 0}_\reg{y}$ is unspecified to allow for different ways to accumulate the result.
So $F':\ket{x,y}\to\ket{x,y\oplus f(x)}$ and $F'':\ket{x,y}\to\ket{x+y \mod 2^k}$ are merely considered to be different implementations of the same quantum function.

\subsubsection{Scratch Space Management}

While quantum functions can be
used to work around the reversible nature of QC, the necessity
to keep a copy of the argument is a problem, as longer
computations will leave registers filled with intermediate results.

Let $F$ be a quantum function with the argument register $\reg{x}$
({\tt quconst}), the target register $\reg{y}$ ({\tt quvoid}) 
and the scratch register $\reg{s}$ ({\tt quscratch})
\begin{equation}
  F(\reg{x},\reg{y},\reg{s}):
  \ket{i}_{\reg{x}}\ket{0}_{\reg{y}}\ket{0}_{\reg{s}}\to
  \ket{i}_{\reg{x}}\ket{f(i)}_{\reg{y}}\ket{j(i)}_{\reg{s}}
\end{equation}
$F$ fills the register $\reg{s}$ with the temporary junk bits $j(i)$.
To reclaim $\reg{s}$, QCL transparently allocates an auxiliary 
register $\reg{t}$ and translates $F$ into an operator $F'$ which 
is defined as \cite{bennett1}
\begin{equation}
  F'(\reg{x},\reg{y},\reg{s},\reg{t})=
  \adj{F}(\reg{x},\reg{t},\reg{s})\,
  \mathtt{fanout}(\reg{t},\reg{y})\,
  F(\reg{x},\reg{t},\reg{s})
\end{equation}
The {\it fanout} operator is a quantum function defined as
\begin{equation}
  \mathtt{fanout}: \ket{i}\ket{0}\to\ket{i}\ket{i}
\end{equation}
The application of $F'$ restores the scratch register 
$\reg{s}$ and the auxiliary register $\reg{t}$ to $\ket{0}$
while preserving the function value in the target register 
$\reg{y}$: 
\begin{equation}
  \ket{i,0,0,0}\to\ket{i,0,j(i),f(i)}\to
  \ket{i,f(i),j(i),f(i)}\to\ket{i,f(i),0,0}
\end{equation}

\subsection{Quantum Conditions}\lab{qucond}

\subsubsection{Conditional Operators}

Classical programs allow the conditional execution of
code in dependence on the content of a boolean variable
(conditional branching).

A unitary operator, on the other hand, is static and
has no internal flow-control. 
Nevertheless, we can conditionally apply an $n$-qubit operator 
$U$ to a quantum register by using an {\it enable} 
qubit and define an $n+1$ qubit operator $U'$
\begin{equation}
  U'=\matr{cc}{I(n) & 0 \\ 0 & U}
\end{equation}
So $U$ is only applied to base-vectors where the enable bit
is set.
This can be easily extended to enable-registers of arbitrary length.

\begin{definition}
A conditional operator $U_{[[\reg{e}]]}$ with the enable register
$\reg{e}$ is a unitary operator of the form
{$$
  U_{[[\reg{e}]]}:\ket{i,\epsilon}=\ket{i}\ket{\epsilon}_\reg{e} 
  \to \alt{ 
  (U\,\ket{i})\,\ket{\epsilon}_\reg{e} &
  \;\mbox{if}\; \epsilon=111\ldots \\
  \ket{i}\ket{\epsilon}_\reg{e} & \;\mbox{otherwise} 
  }
$$}
\end{definition}

A conditional version of the increment operator from \ref{inc}
can be explicitly implemented as
\bqcl
\begin{verbatim}
qufunct cinc(qureg x,quconst e) {
  int i;
  for i = #x-1 to 1 step -1 { CNot(x[i],x[0:i-1] & e); }
  CNot(x[0],e);
}
\end{verbatim}
\eqcl

QCL can automatically derive $U_{[[\reg{e}]]}$ for an operator $U$ if its declaration is prefixed with {\tt cond}.
\bqcl
\begin{verbatim}
cond qufunct inc(qureg x,quconst e) { ... }
\end{verbatim}
\eqcl
The enable register can be set by a quantum {\tt if}-statement and QCL
will transparently transform the defined operator into its
conditional version when necessary.

\subsubsection{Quantum if-statement}

Just like the concept of quantum functions allows the computation
of irreversible boolean functions with unitary operators, conditional
operators allow conditional branching depending on unobservable qubits.

Given the above definitions, the statement 
{\tt\verb=if e { inc(x); }=} 
is equivalent to the explicit call of {\tt cinc(x,e)} and
{\tt\verb=if e { inc(x); } else { !inc(x); }=} is equivalent to the sequence
\bqcl
\begin{verbatim}
cinc(x,e);      // conditional increment
Not(e);         // invert enable qubit
!cinc(x,e);     // conditional decrement
Not(e);         // restore enable qubit
\end{verbatim}
\eqcl

Quantum if-statements can be nested. Since operators within the
if- and else-branches are transformed into their conditional 
versions, they must be declared {\tt cond} and must not operate
on any qubits used in the condition.

\subsubsection{Complex Conditions}

Conditions in quantum if-statements are not restricted to
single qubits, but can contain any boolean expression and
also allow the mixing of classical and quantum bits.

\bqcl
\begin{verbatim}
qcl> qureg q[4]; qureg b[1]; qureg a[1];
qcl> H(a & b);  // prepare test-state
[6/32] 0.5 |000000> + 0.5 |010000> + 0.5 |100000> + 0.5 |110000>
qcl> if a and b { inc(q); }
[6/32] 0.5 |000000> + 0.5 |010000> + 0.5 |100000> + 0.5 |110001>
qcl> if a or b { inc(q); }
[6/32] 0.5 |000000> + 0.5 |010001> + 0.5 |100001> + 0.5 |110010>
qcl> if not (a or b) { inc(q); }
[6/32] 0.5 |000001> + 0.5 |010001> + 0.5 |100001> + 0.5 |110010>
\end{verbatim}
\eqcl

QCL produces a sequence of {\tt CNot}-gates to evaluate a {\it quantum condition}.\footnote{Internally, this is achieved by transforming a quantum condition into its exclusive disjunctive normal form \cite{bani}.} If necessary, scratch qubits are transparently allocated and uncomputed again.

\subsubsection{Forking if-statement}

If the body of a quantum if-statement contains statements which
change the program state (e.g. assignments to local
variables), then subsequent operator calls may differ, depending
on whether the if- or the else-branch has been executed.

In that case, QCL follows all possible classical paths throughout
the operator definition (forking), accumulates the conditions of all
visited quantum if-statements and serializes the generated sequence
of operators.

\bqcl
\begin{verbatim}
cond qufunct demux(quconst s,qureg q) {
  int i;
  int n = 0;
  for i=0 to #s-1 {             // accumulate content of
    if s[i] { n=n+2^i; }        //   selection register in a
  }                             //   classical variable
  Not(q[n]);                    // flip selected output qubit
}
\end{verbatim}
\eqcl

Figure \ref{demux} shows the quantum circuit \cite{nielsen} generated by {\tt demux(s,q)} in the case of a $2$-qubit selection register $\reg{s}$.

\graphic{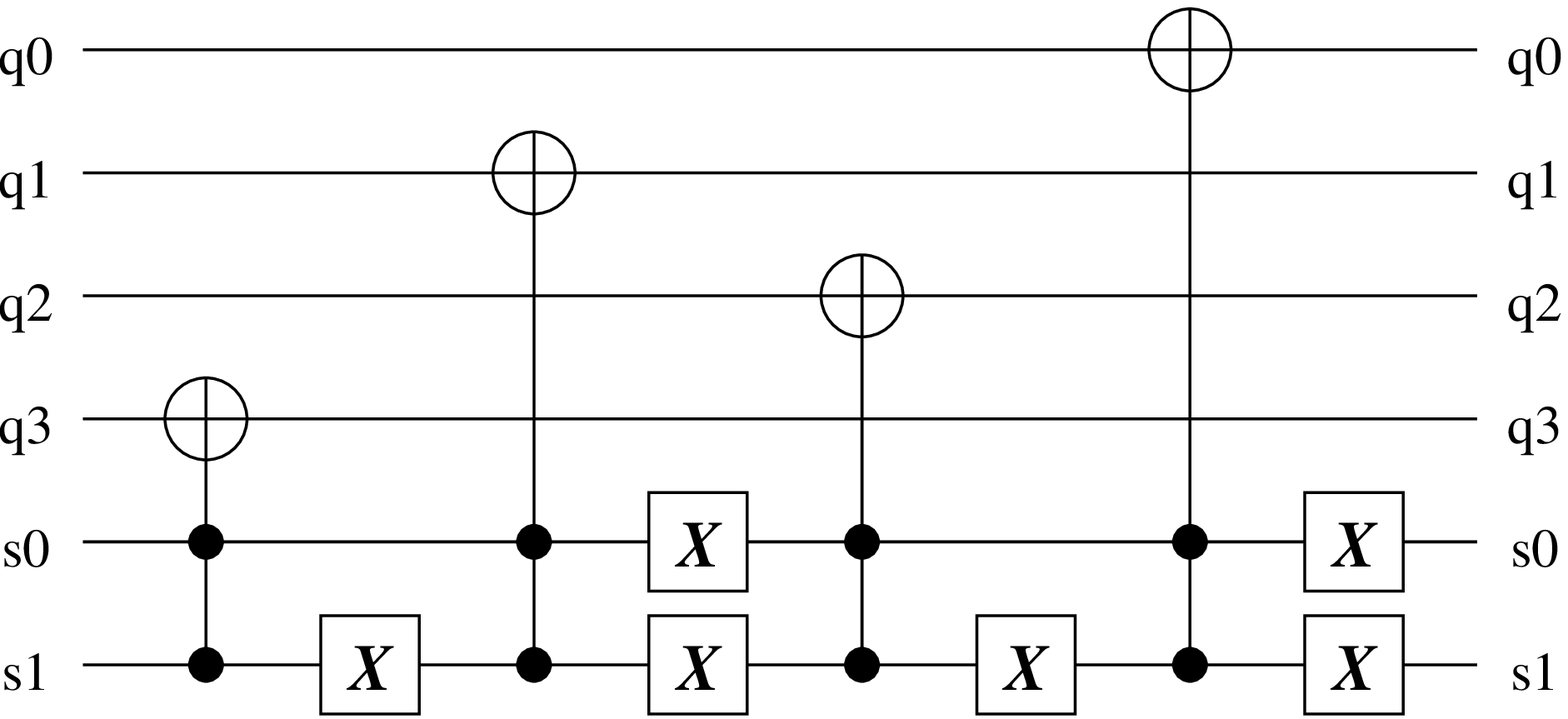}{70mm}{a quantum demultiplexer}

Forking if-statements may only appear within an operator definition to
assure that the different execution threads can be joined again.

\section{Conclusions}

Throughout the history of classical computing, hardware development has always been accompanied by a corresponding improvement of programming methodology. 
The formalism of quantum circuits --- the moral equivalent to hand written machine code --- seems inadequate for quantum computers with more than a couple of qubits, so more abstract methods will eventually be required.

We have demonstrated how well established concepts of classical programming languages like subroutines, local variables or conditional branching can be ported to the field of quantum computing.
Besides providing a new level of abstraction, we also hope that a quantum programming language which semantically integrates those concepts will allow for a better and more intuitive understanding of non-classical algorithms.

\end{document}